\documentclass[prd,aps,showkeys,superscriptaddress,two column]{revtex4-1}
\usepackage{amsmath}
\usepackage{amssymb}
\usepackage{amsthm}
\usepackage{mathrsfs}
\usepackage{graphicx}
\usepackage{fancyhdr}
\usepackage{array}
\usepackage{simplewick}
\usepackage{latexsym}
\usepackage[all]{xy}
\usepackage{eufrak}
\usepackage{euscript}
\usepackage{enumerate}
\usepackage{dsfont}
\usepackage{slashed}
\usepackage{hyperref}
\usepackage{caption}

\begin{document}

%%%%%%%%%%%%%%%%%%%%%%%%%%%%%%%%%%%%%%%%%%%%%%%%%%%%%%%%%%%%%

\title{The Chern-Simons term in a dual Josephson junction}

\author{L. S. Grigorio}
\email{leogrigorio@gmail.com}
\affiliation{Centro Federal de Educa\c{c}\~ao Tecnol\'ogica Celso Suckow da Fonseca, 28635-000, Nova Friburgo, Brazil}

\author{M. S. Guimaraes}
\email{msguimaraes@uerj.br}
\affiliation{Instituto de F\'isica, Universidade do Estado do Rio de Janeiro, 20550-013, Rio de Janeiro, Brazil}

\author{R. Rougemont}
\email{romulo@if.ufrj.br}
\affiliation{Instituto de F\'isica, Universidade Federal do Rio de Janeiro, 21941-972, Rio de Janeiro, Brazil}

\author{C. Wotzasek}
\email{clovis@if.ufrj.br}
\affiliation{Instituto de F\'isica, Universidade Federal do Rio de Janeiro, 21941-972, Rio de Janeiro, Brazil}

\author{C. A. D. Zarro}
\email{carlos.zarro@if.ufrj.br}
\affiliation{Instituto de F\'isica, Universidade Federal do Rio de Janeiro, 21941-972, Rio de Janeiro, Brazil}

%\date{\today}

\begin{abstract}
A dual Josephson junction corresponding to a $(2+1)$-dimensional non-superconducting layer sandwiched between two $(3+1)$-dimensional dual superconducting regions constitutes a model of localization of a $U(1)$ gauge field within the layer. Monopole tunneling currents flow from one dual superconducting region to another due to a phase difference between the wave functions of the monopole condensate below and above the non-superconducting layer when there is an electromagnetic field within the layer. These magnetic currents appear within the $(2+1)$-dimensional layer as a gas of magnetic instanton events and a weak electric charge confinement is expected to take place at very long distances within the layer. In the present work, we consider what happens when one introduces fermions in this physical scenario. Due to the dual Meissner effect featured in the dual superconducting bulk, it is argued that unconfined fermions would be localized within the $(2+1)$-dimensional layer, where their quantum fluctuations radiatively induce a Chern-Simons term, which is known to destroy the electric charge confinement and to promote the confinement of the magnetic instantons.
\end{abstract}

%\pacs{Valid PACS appear here}
% PACS, the Physics and Astronomy Classification Scheme.
% Valid PACS numbers may be entered using the \verb+\pacs{#1} command.

\keywords{Chern-Simons term, instantons, confinement, dual supercondutor, Josephson junction.}
% Use showkeys class option if keyword display desired

%%%%%%%%%%%%%%%%%%%%%%%%%%%%%%%%%%%%%%%%%%%%%%%%%%%%%%%%%%%%%

\maketitle

\section{Introduction}
\label{sec1}

In the present work, we propose an extension of the model presented in \cite{tetradis}. There, it was considered a dual Josephson junction comprising a $(2+1)$-dimensional non-superconducting layer sandwiched between two $(3+1)$-dimensional dual superconducting regions, which correspond to a condensate of magnetic monopoles \cite{ripka}. At the tree level, this physical setup localizes an effectively free electromagnetic gauge potential in the $(2+1)$-dimensional layer, since the massless photons within the layer cannot escape from it to propagate far into the interior of the dual superconducting bulk due to the mass acquired by gauge bosons within such a magnetic condensate medium. However, when quantum effects are taken into account, monopole tunneling currents can flow from one dual superconducting region to another due to a phase difference between the wave functions of the monopole condensate below and above the layer when there is an electromagnetic field within the layer. This is the dual Josephson effect, whose aftermath is to provide a small mass to the photons, generating a weak dual Meissner effect and an associate weak confinement of electric charges within the layer. This can be understood from the fact that the monopole tunneling currents appear within the $(2+1)$-dimensional layer as a gas of magnetic instantons, which are known to cause the confinement of electric charges \cite{polyakov}. Therefore, when quantum effects are considered, the effective model for the $(2+1)$-dimensional layer also displays electric charge confinement, although only at large length scales, which is the reason why this is said to be a weak electric charge confinement \cite{tetradis}. We shall briefly review the main points of this discussion in section \ref{sec2}.

In section \ref{sec3}, we present our original results by considering what happens when dynamical fermions are included in the above physical scenario. We argue that due to the dual Meissner effect, fermions would be confined in mesonic fermion-antifermion pairs within the dual superconducting bulk, while unconfined fermions would be localized within the $(2+1)$-dimensional layer, where their quantum fluctuations radiatively induce a Chern-Simons (CS) term \cite{mcs}, which is known to destroy the electric charge confinement \cite{affleck} and to drive the confinement of the magnetic instantons \cite{pisarski,diamantini,mcsmon,artigao}. The induction of the CS term in $(2+1)$-dimensions can also be accomplished by interpreting the fermion quantum fluctuations as an electric condensation process \cite{artigao,santiago} through the use of the so-called generalized Julia-Toulouse approach for condensation of topological currents.

The Julia-Toulouse approach is originally \cite{jt,qt} a prescription used to construct a low energy effective field theory for a system characterized by a condensate of topological currents, by starting from the model describing the system in the regime where these currents are dilutely distributed through space and by taking into account the symmetries expected for the condensed regime. In \cite{dafdc} we showed that the original Julia-Toulouse prescription \cite{jt,qt} is the electromagnetic dual, in a certain limit, of the formalism developed in \cite{banks,mvf}, where the condensation of topological currents is controlled by an adequate fugacity term inserted into a partition function with an associate ensemble of defects. In \cite{artigao,mbf-proca-m} we unified and extended both prescriptions into a generalized formalism, whose main feature, in the condensed phase, consists in a careful treatment of a local symmetry which we call as the Dirac brane symmetry, which is independent of the usual gauge symmetry \cite{mvf} and corresponds to the freedom of deforming Dirac strings \cite{dirac} without any observable consequences. We call this generalized prescription as the generalized Julia-Toulouse approach (GJTA) \cite{artigao,mbf-proca-m}.

We shall apply the GJTA in section \ref{sec3} to consistently induce the CS term in the $(2+1)$-dimensional layer in the presence of external magnetic instantons associated to the dual Josephson tunneling currents. As the result, the weak electric charge confinement featured within the layer in the absence of fermions is destroyed when fermions are present and the magnetic instantons become confined in this setup due to the presence of an electric condensate effectively described by the CS term.

Throughout this paper, we are going to make use of natural units with $c=\hbar=1$.

\section{Dual Josephson junction in the absence of fermions and weak electric charge confinement in the $(2+1)$-dimensional layer}
\label{sec2}

We begin by giving a brief review of the main points involved in the physical scenario considered in \cite{tetradis}. The effective tree level action describing the dual Josephson junction proposed in \cite{tetradis} is given by:
\begin{align}
S_{\textrm{DJJ}}[C_\mu,\phi]&=\int_{\mathbb{R}^{1,3}}d^4x\left[-\frac{1}{4}\left(\partial_{[\mu}C_{\nu]}\right)^2 +\frac{1}{2}|(\partial_\mu+igC_\mu)\phi|^2\right.\nonumber\\
&\left.-\frac{\lambda}{4}\left(|\phi|^2-v^2\right)^2\right],
\label{eq1}
\end{align}
where $C_\mu$ is the dual gauge potential \cite{ripka,mvf}, $g$ is the monopole charge and the wave function of the magnetic condensate is given by:
\begin{align}
\phi(x)=\begin{cases}
        \rho(x)e^{i\theta(x)} & \text{for $\,\,\,\,x \in \mathbb{R}^{1,3}\backslash\,\mathbb{R}^{1,2}_{(|z|<\varepsilon/2)}$},\\
        0                     & \text{for $\,\,\,\,x \in \mathbb{R}^{1,2}_{(|z|<\varepsilon/2)}\subset\mathbb{R}^{1,3}$},
          \end{cases}
\label{eq2}
\end{align}
where $\mathbb{R}^{1,3}\backslash\,\mathbb{R}^{1,2}_{(|z|<\varepsilon/2)}$ corresponds to the dual superconducting bulk and the $(2+1)$-dimensional non-superconducting layer is placed at $x^3\equiv z=0$, having a small width $\varepsilon$. The dual superconducting bulk described by the dual Abelian Higgs model features a dual Meissner effect which confines electric charges, as usual. At the tree level, we see from (\ref{eq1}) and $(\ref{eq2})$ that the monopole condensate field $\phi(x)$ vanishes within the $(2+1)$-dimensional layer and, therefore, we have a massless photon freely propagating within the layer, at least classically. In fact, due to the boundary conditions set by the dual superconductor at low energies, the only components of the observable electromagnetic fields that survive within the layer are \cite{tetradis}:
\begin{align}
E_x=-\frac{\partial C^3}{\partial y},\,\,\,E_y=\frac{\partial C^3}{\partial x},\,\,\,B_z=-\frac{\partial C^3}{\partial t}.
\label{eq3}
\end{align}
This allows one to suitably define an effective $(2+1)$-dimensional real scalar field $\varphi(x,y,t):=C^3(x,y,z=0,t)$ and write down the following effective field theory describing the tree level physics within the $(2+1)$-dimensional layer:
\begin{align}
S_{\textrm{eff}}[\varphi]=\int_{\mathbb{R}^{1,2}}d^3x\,\frac{1}{2}\left(\partial_\mu\varphi\right)^2,
\label{eq4}
\end{align}
which is the electromagnetic dual of a Maxwell action in $(2+1)$-dimensions \cite{artigao,mvf}:
\begin{align}
\tilde{S}_{\textrm{eff}}[a_\mu]=\int_{\mathbb{R}^{1,2}}d^3x\left[-\frac{1}{4}\left(\partial_{[\mu}a_{\nu]}\right)^2\right].
\label{eq5}
\end{align}

Let us now see how quantum effects modify this classical picture within the layer. When we consider a phase difference between the monopole condensate wave functions below and above the non-superconducting layer, monopole currents tunnel across the layer, appearing within it as a gas of magnetic instanton events. In the dual superconducting bulk, a monopole current $j^\mu$ couples minimally to the dual gauge potential $C_\mu$ and when it crosses the layer at $z=0$ it is oriented along the $z$-direction. Therefore, this coupling appears within the layer as $gC_\mu j^\mu\stackrel{\textrm{\tiny{layer}}}{\longrightarrow} g\varphi j^3$ and the presence of the magnetic instanton gas is accounted for by the following partition function:
\begin{align}
Z=\int\mathcal{D}\varphi\exp\left\{i\int_{\mathbb{R}^{1,2}}d^3x\,\frac{1}{2}\left(\partial_\mu\varphi\right)^2\right\} Z_I[\varphi],
\label{eq6}
\end{align}
where:
\begin{align}
Z_I[\varphi]=\sum_{\left\{j^3\right\}}\exp\left\{ig\int_{\mathbb{R}^{1,2}}d^3x\,\varphi j^3\right\},
\label{eq7}
\end{align}
encodes the magnetic instanton gas contribution. Let us explicitly evaluate this contribution by considering the dilute gas approximation \cite{polyakov}.

In what follows, we shall take basically the same steps presented in \cite{jt-schwinger}. We begin by wick-rotating (\ref{eq7}) to the euclidean space $\mathbb{R}^3$ ($t\mapsto -it_E, d^3x\mapsto -id^3x_E,j^3\mapsto -ij^3_E,\phi\mapsto \phi_E$):
\begin{align}
Z_I^E[\varphi_E]=\sum_{\left\{j^3_E\right\}}\exp\left\{-ig\int_{\mathbb{R}^3}d^3x_E\,\varphi_E j^3_E\right\}.
\label{eq8}
\end{align}
Firstly, we take the contribution of a single instanton with winding number $+1$ into consideration:
\begin{align}
\int_{\mathbb{R}^3} d^3x^\beta_E\, \eta&\exp\left\{-ig\int_{\mathbb{R}^3} d^3x_E\, \varphi_E(x_E)\delta^{(3)}(x_E-x_E^\beta)\right\}=\nonumber\\
&=\int_{\mathbb{R}^3} d^3x^\beta_E\, \eta\exp\left\{-ig\varphi_E(x^\beta_E)\right\},
\label{eq9}
\end{align}
where $\eta$ is the instanton fugacity corresponding to the probability density of existence of a single instanton with winding number $+1$. Notice we are integrating over all possible instanton locations $x_E^\beta$. Summing over all possible system configurations with an arbitrary number of instantons with winding number $+1$ and assuming that they do not interact among themselves, we get:
\begin{align}
\sum_{N_+=0}^\infty&\frac{1}{N_+!}\left(\int_{\mathbb{R}^3} d^3x^\beta_E\, \eta \exp\left\{-ig\varphi_E(x^\beta_E)\right\}\right)^{N_+}=\nonumber\\
&=\exp\left\{\int_{\mathbb{R}^3} d^3x_E\, \eta e^{-ig\varphi_E(x_E)}\right\},
\label{eq10}
\end{align}
where the factor $(N_+!)^{-1}$ is needed to account for the fact that we have indistinguishable instantons. Considering now the contribution of an arbitrary number of antinstantons with winding number $-1$ and assuming a small instanton fugacity $\eta$, such that the contribution of instantons and antinstantons with higher winding numbers becomes exponentially suppressed in the partition function $Z_I^E[\varphi_E]$, we obtain the approximation:
\begin{align}
Z_I^E[\varphi_E]\approx&\exp\left\{\int_{\mathbb{R}^3} d^3x_E\, \eta e^{-ig\varphi_E(x_E)}\right\}\times\nonumber\\
&\times\exp\left\{\int_{\mathbb{R}^3} d^3x_E\, \eta e^{+ig\varphi_E(x_E)}\right\}=\nonumber\\
&=\exp\left\{\int_{\mathbb{R}^3} d^3x_E\, 2\eta\cos(g\varphi_E(x_E))\right\}.
\label{eq11}
\end{align}
Therefore, the net effect of a gas of instantons consists in generating a cosine self-interaction term for the scalar field \cite{polyakov}. Wick-rotating (\ref{eq11}) back to the Minkowski space $\mathbb{R}^{1,2}$ and substituting the result in (\ref{eq6}), we get the Sine-Gordon model as the effective field theory within the $(2+1)$-dimensional layer when the dual Josephson currents are taken into account:
\begin{align}
Z\approx\int\mathcal{D}\varphi\exp\left\{i\int_{\mathbb{R}^{1,2}}d^3x\left[\frac{1}{2}\left(\partial_\mu\varphi\right)^2
+2\eta\cos(g\varphi)\right]\right\}.
\label{eq12}
\end{align}
From the above equation we conclude that due to the monopole tunneling currents the scalar field acquires an effective mass $m^2_{\textrm{eff}}=2\eta g^2$. This is a small mass, since in the derivation of the Sine-Gordon model (\ref{eq12}) we assumed a small instanton fugacity $\eta$.

We see from (\ref{eq12}) that the next-to-leading order effective action describing the physics within the $(2+1)$-dimensional layer of the dual Josephson junction setup is given by:
\begin{align}
S_{\textrm{eff}}^{\textrm{NLO}}[\varphi]\approx\int_{\mathbb{R}^{1,2}}d^3x\left[\frac{1}{2}\left(\partial_\mu\varphi\right)^2
-\frac{m^2_{\textrm{eff}}}{2}\varphi^2\right].
\label{eq13}
\end{align}
If we now insert two opposite electric charges into this system via a non-minimal coupling with the scalar field, it can be shown that they interact via an effective potential that features two parts: a short-range Yukawa interaction plus a confining term that is linear in the intercharges separation and proportional to $m^2_{\textrm{eff}}$ \cite{polyakov,artigao}. This picture is qualitatively the same that happens in the $(3+1)$-dimensional dual superconducting bulk, but quantitatively, the electric confinement phenomenon observed within the $(2+1)$-dimensional layer only becomes noticeable at much larger distances than its bulk counterpart, since $m^2_{\textrm{eff}}$ is much smaller than the mass acquired by the dual gauge field $C_\mu$ in the bulk due to the dual Higgs mechanism. In this sense, it is said that the electric charge confinement featured within the layer is weak \cite{tetradis}.

It is important to mention that the same physics just discussed could be equivalently approached in the dual picture, in terms of the massless gauge potential $a_\mu$ appearing in (\ref{eq5}), instead of its electromagnetic dual, the massless scalar field $\varphi$ featured in (\ref{eq4}) \cite{polyakov,artigao}. The magnetic instantons would couple non-minimally to the $a_\mu$ field and their condensation would imply in the rank-jump phenomenon, with the confinement of electric charges being described in terms of a massive Kalb-Ramond field, the electromagnetic dual in $(2+1)$-dimensions of the massive scalar field featured in (\ref{eq12}) and (\ref{eq13}). For the interested reader, we refer the section IV of reference \cite{artigao}, where these issues are discussed in details.

\section{Fermions, the induction of the Chern-Simons term and magnetic instanton confinement in the $(2+1)$-dimensional layer}
\label{sec3}

Let us now discuss how the presence of fermions in the system drastically changes the physics within the $(2+1)$-dimensional layer reviewed in the last section.

In what follows, we shall work in terms of the gauge field $a_\mu$ present in (\ref{eq5}), which is the electromagnetic dual of the scalar field $\varphi$ within the layer. The reason for this is because while fermions in the bulk couple minimally to the $(3+1)$-dimensional gauge potential $A_\mu$, its coupling to the dual gauge potential $C_\mu$ is not well understood. Since the scalar field $\varphi$ is the inheritance of the $C_\mu$ field within the $(2+1)$-dimensional layer, we should find the same problem in trying to define the coupling between fermions and $\varphi$. However, since the $a_\mu$ field is just the restriction of the bulk gauge field $A_\mu$ within the layer, it couples minimally to fermions localized within the layer. In this way, we write down the following effective field theory within the $(2+1)$-dimensional layer, describing the interaction of the $a_\mu$ field minimally coupled to fermions and non-minimally coupled to external magnetic instantons:
\begin{align}
Z_\psi[j^3]=\int_{\textrm{G.F.}}\mathcal{D} & a_\mu \int\mathcal{D}\bar{\psi}\mathcal{D}\psi\exp\left\{i\int_{\mathbb{R}^{1,2}}d^3x
\left[-\frac{1}{4}\left(\partial_{[\mu}a_{\nu]}\right.\right.\right.\nonumber\\
&\left.\left.\left.-\,g\tilde{\chi}_{\mu\nu}\right)^2+\bar{\psi}\left(i\slashed{D}-m_\psi\right)\psi\right]\right\},
\label{eq14}
\end{align}
where $j^3=\frac{1}{2}\varepsilon^{3\mu\nu\beta}\partial_\mu\tilde{\chi}_{\nu\beta}\equiv \frac{1}{2}\epsilon^{\mu\nu\beta} \partial_\mu\tilde{\chi}_{\nu\beta}$ is the scalar instanton current that localizes a magnetic instanton event within the layer, $\tilde{\chi}_{\mu\nu}=\epsilon_{\mu\nu\beta}\chi^\beta$ is the associated Chern-Kernel, which localizes the magnetic Dirac string that has the magnetic instanton event in its boundary, $\slashed{D}=\gamma^\mu(\partial_\mu+iea_\mu)$ is the $(2+1)$-dimensional covariant Dirac operator, $e$ is the electric fermion charge and the acronym ``G.F.'' in the functional measure stands for an arbitrary gauge fixing procedure that must be adopted at some stage of the calculations, in order to fix the gauge ambiguity present in (\ref{eq14}).

The $(3+1)$-dimensional dual superconducting bulk features a dual Meissner effect which tends to expel all the electric fields from the interior of the bulk, causing the confinement of electric charges \cite{ripka}. Dynamical fermions within the dual superconducting bulk always appear in mesonic bound states with vanishing electric charge and if a fermion-antifermion pair is elongated beyond a certain critical distance, it becomes energetically favorable to create a new fermion-antifermion pair and, therefore, a string breaking process occurs. In this way, unconfined fermions can only lie within the $(2+1)$-dimensional non-superconducting layer of the dual Josephson junction setup.

In the absence of instantons, the integration of the Dirac fields in (\ref{eq14}) leads to a fermion determinant which is responsible for generating a CS term at the one loop level \cite{mcs,dunne}. However, when instantons are present, a direct evaluation of this fermion determinant is not an obvious task because the gauge potential $a_\mu$ is singular over the magnetic Dirac strings \cite{mcsmon}. The strategy we are going to adopt in the sequel to circumvent this difficult is to make use of the GJTA. In the case without instantons, it was shown in \cite{artigao,santiago} that the GJTA can reproduce the effect of the one loop fermion fluctuations, namely, the induction of a CS term, by interpreting this term as arising due to a condensation of classical electric charges that breaks parity and time reversal symmetries. These classical electric charges are represented by electric world-lines instead of fermion fields. Due to this equivalence between the two methods (one loop fermion determinant calculation and electric world-lines condensation via GJTA), we applied in \cite{mcsmon,artigao} the GJTA to induce the CS term in the presence of instantons. As the result, the external magnetic instantons become confined in the Maxwell-Chern-Simons (MCS) theory due to the electric condensate effectively described by the CS term. Similar conclusions for magnetic instantons in the MCS model were firstly reached in references \cite{affleck,pisarski,diamantini,ht}, although the CS term was not interpreted there as arising due to an electric condensation process.

Let us now proceed via GJTA to obtain the MCS theory in the presence of magnetic instantons as a result of an electric condensation process within the $(2+1)$-dimensional layer. As explained above, within the GJTA we shall substitute the fermion description of the electric charges by a classical description in terms of electric world-lines. The quantum effect of integrating over the fermions at the one loop level will be equivalently reproduced within the GJTA by considering a condensation of these electric world-lines at low energies. The first step in our route, therefore, consists in writing down an effective theory for the classical electric charges in a regime where they are dilutely distributed through space, also taking into account the presence of the external instantons due to the dual Josephson tunneling currents:
\begin{align}
Z_{\textrm{dil}}[j^3,J_\mu]=\int_{\textrm{G.F.}}\mathcal{D}a_\mu & \exp\left\{i\int_{\mathbb{R}^{1,2}}d^3x
\left[-\frac{1}{4}\left(\partial_{[\mu}a_{\nu]}\right.\right.\right.\nonumber\\
&\left.\left.\left.-\,g\tilde{\chi}_{\mu\nu}\right)^2+ea_\mu J^\mu\right]\right\},
\label{eq15}
\end{align}
where $J_\mu=\epsilon_{\mu\nu\beta}\partial^\mu\tilde{\Sigma}^\beta$ is the electric current localizing the electric world-line, which is the boundary of the Chern-Kernel $\tilde{\Sigma}_\mu=\frac{1}{2}\epsilon_{\mu\nu\beta}\Sigma^{\nu\beta}$ localizing the electric Dirac brane corresponding to the world-surface of the electric Dirac string attached to the electric charge $e$.

Besides the usual gauge ambiguity, $a_\mu\mapsto a_\mu+\partial_\mu\omega$, which implies the electric charge conservation, $\partial_\mu J^\mu=0$, there are also two extra local symmetries featured in (\ref{eq15}) \cite{artigao,mvf}. The first one is the electric Dirac brane symmetry corresponding to the invariance of (\ref{eq15}) under deformations of the electric Dirac branes keeping fixed their physical boundaries corresponding to the electric world-lines:
\begin{align}
\tilde{\Sigma}_\mu\mapsto\tilde{\Sigma}_\mu+\partial_\mu\tilde{\xi},
\label{eq16}
\end{align}
where $\tilde{\xi}=\frac{1}{3!}\epsilon^{\mu\nu\beta}\xi_{\mu\nu\beta}$ is a delta distribution localizing the volume spanned in $\mathbb{R}^{1,2}$ by the deformation of the Dirac surface localized by $\tilde{\Sigma}_\mu$, keeping fixed its boundary. The another extra local symmetry is the magnetic Dirac brane symmetry corresponding to the invariance of (\ref{eq15}) under deformations of the magnetic Dirac strings keeping fixed their physical boundaries corresponding to the magnetic instanton events:
\begin{align}
\tilde{\chi}_{\mu\nu}&\mapsto\tilde{\chi}_{\mu\nu}+\partial_{[\mu}\tilde{\sigma}_{\nu]},\nonumber\\
a_\mu&\mapsto a_\mu+g\tilde{\sigma}_{\mu},
\label{eq17}
\end{align}
where $\tilde{\sigma}_{\mu}=\frac{1}{2}\epsilon_{\mu\nu\beta}\sigma^{\nu\beta}$ is a delta distribution localizing the surface spanned in $\mathbb{R}^{1,2}$ by the deformation of the Dirac string localized by $\tilde{\chi}_{\mu\nu}$, keeping fixed its boundary. Notice that the gauge potential $a_\mu$ is dislocated under the magnetic Dirac brane transformation (\ref{eq17}) because it is singular over the magnetic Dirac branes \cite{artigao,mvf}. In fact, not only $a_\mu$, but also its exterior derivative $\partial_{[\mu}a_{\nu]}$ are singular over the magnetic Dirac branes. However, the singularity present in $\partial_{[\mu}a_{\nu]}$ is exactly canceled out by the singular Dirac string term, such that the non-minimal coupling $\left(\partial_{[\mu}a_{\nu]}-g\tilde{\chi}_{\mu\nu}\right)$ is the regular structure representing the observable electromagnetic fields in the presence of magnetic defects \cite{ripka,artigao,mvf}.

While the electric Dirac brane symmetry (\ref{eq16}) is trivially realized in the partition function (\ref{eq15}), the magnetic Dirac brane transformation (\ref{eq17}) is not an immediately obvious symmetry of (\ref{eq15}) due to the minimal coupling term, and its analysis has some very important consequences. In fact, notice that under the transformation (\ref{eq17}), while the non-minimal coupling in (\ref{eq15}) is clearly invariant, the minimal coupling is shifted by an intersection number between $J^\mu$ and $\tilde{\sigma}_\mu$ times the product $eg$. Therefore, in order to the unphysical magnetic Dirac strings to remain unobservable at the quantum level in the partition function (\ref{eq15}), one must require as a consistency condition that the product between the electric and magnetic charges is quantized in integer multiples of $2\pi$, which is the celebrated Dirac charge quantization condition \cite{dirac}: $eg=2\pi n,\,n\in\mathbb{Z}$. However, although necessary, the Dirac charge quantization condition is not sufficient to guarantee the invariance of the minimal coupling term in (\ref{eq15}) under the magnetic Dirac brane transformation (\ref{eq17}). Indeed, since the minimal coupling term has support over the electric world-lines, in order to the action of the system to remain regular over the whole space, one must require as a further consistency condition that the magnetic Dirac strings do not touch the electric world-lines, otherwise at the points where such crossings happen, the minimal coupling would diverge, since the $a_\mu$ field is singular over the magnetic Dirac strings. This restriction is known as the Dirac's veto \cite{dirac}. The Dirac's veto together with the Dirac charge quantization condition are two necessary and sufficient consistency conditions that guarantee the invariance of the theory under the magnetic Dirac brane transformations, as detailed discussed in \cite{mbf-proca-m}. Consequently, we can precisely state the magnetic Dirac brane symmetry as corresponding to the local invariance of the theory under the deformations (\ref{eq17}), provided the Dirac charge quantization condition and the Dirac's veto are both satisfied: we are allowed to move the magnetic Dirac strings to any place not occupied by the electric world-lines \cite{mbf-proca-m}.

Having written the partition function (\ref{eq15}) for the electrically diluted regime within the layer and having analyzed its local symmetries, we are now prepared to proceed with the GJTA. As discussed before, we are interested in obtaining the MCS theory in the presence of external instantons as a result of an electric condensation process, emulating the effect of fermion quantum fluctuations within the layer. The electric condensation process associated to the emergence of the CS term, $a_\mu\epsilon^{\mu\nu\beta}\partial_\nu a_\beta$, must break the discrete parity and time reversal symmetries. With this in mind, our next step consists in defining the partition function for the electrically condensed regime by adding to the partition function (\ref{eq15}) a fugacity term allowing the proliferation of the electric world-lines until they establish a macroscopically continuous medium corresponding to the electric condensate. More precisely, we shall consider a proliferation of the electric Dirac branes localized by $\tilde{\Sigma}_\mu$ and this, in turn, will imply the proliferation of the electric world-lines that live in their boundaries. In this way, the form of the fugacity term to be added to (\ref{eq15}) is determined by taking a derivative expansion of the electric Dirac branes $\tilde{\Sigma}_\mu$ compatible with the desired symmetry content and retaining only the term with lowest order in derivatives, which is the term giving the dominant contribution for the low energy physics. Regarding the symmetry content we are interested in, the term to be added must preserve all the local symmetries of the system, in consonance with Elitzur's theorem \cite{elitzur}, and must also preserve the Lorentz symmetry and break parity and time reversal symmetries, as discussed before. The term with lowest order in derivatives satisfying all these requirements is of the form $\tilde{\Sigma}_\mu\epsilon^{\mu\nu\beta}\partial_\nu\tilde{\Sigma}_\beta$. Therefore, we write down the following partition function for the electrically condensed regime:
\begin{align}
&Z_{\textrm{cond}}[j^3]=\sum_{\left\{\tilde{\Sigma}_\mu\right\}}\int_{\textrm{G.F.}}\mathcal{D}a_\mu \exp\left\{i\int_{\mathbb{R}^{1,2}}d^3x\left[-\frac{1}{4}\left(\partial_{[\mu}a_{\nu]}\right.\right.\right.\nonumber\\
&\left.\left.\left.-\,g\tilde{\chi}_{\mu\nu}\right)^2+ea_\mu\epsilon^{\mu\nu\beta}\partial_\nu\tilde{\Sigma}_\beta +\alpha\tilde{\Sigma}_\mu\epsilon^{\mu\nu\beta}\partial_\nu\tilde{\Sigma}_\beta\right]\right\},
\label{eq18}
\end{align}
where $\alpha$ is an adimensional parameter that will be fixed afterward (in eq. (\ref{eq28})) and we are summing over all electric brane configurations.

We now proceed with some formal manipulations in order to recast the partition function (\ref{eq18}) for the condensed regime in a more suitable form to analyze its physical meaning. We firstly rewrite (\ref{eq18}) as:
\begin{align}
Z_{\textrm{cond}}[j^3]=&\sum_{\left\{\tilde{\Sigma}_\mu\right\}}\int_{\textrm{G.F.}}\mathcal{D}a_\mu\int\mathcal{D}b_\mu\, \delta\left[b_\mu-\tilde{\Sigma}_\mu\right]\nonumber\\
\exp&\left\{i\int_{\mathbb{R}^{1,2}}d^3x\left[-\frac{1}{4}\left(\partial_{[\mu}a_{\nu]}-g\tilde{\chi}_{\mu\nu}\right)^2
\right.\right.\nonumber\\
&\left.\left.+\,ea_\mu\epsilon^{\mu\nu\beta}\partial_\nu b_\beta +\alpha b_\mu\epsilon^{\mu\nu\beta}\partial_\nu b_\beta\right]\right\}.
\label{eq19}
\end{align}
In the next step, we are going to make use of the following version of the generalized Poisson identity (GPI) \cite{dafdc,mvf}:
\begin{align}
\sum_{\left\{\tilde{\Sigma}_\mu\right\}}\delta\left[b_\mu-\tilde{\Sigma}_\mu\right] =\sum_{\left\{\Omega_\mu\right\}}\exp\left\{2\pi i\int_{\mathbb{R}^{1,2}}d^3x\,b_\mu\Omega^\mu\right\},
\label{eq20}
\end{align}
where $\Omega_\mu$ is a delta distribution localizing certain lines in $\mathbb{R}^{1,2}$. As detailed discussed in Appendix A of \cite{dafdc}, the GPI works a geometric analogue of the Fourier transform: when the ensemble of branes on one side of (\ref{eq20}) becomes prolific, the ensemble of branes of complementary dimension on the other side of the GPI becomes diluted. From this, we conclude that the lines localized by $\Omega_\mu$ must be interpreted as vortex lines over the electric condensate: their proliferation (dilution) implies the dilution (proliferation) of the electric world-lines that lie on the boundaries of the electric Dirac branes $\tilde{\Sigma}_\mu$.

Using the GPI (\ref{eq20}), we rewrite (\ref{eq19}) as:
\begin{align}
Z_{\textrm{cond}}[j^3]&=\sum_{\left\{\Omega_\mu\right\}}\int_{\textrm{G.F.}}\mathcal{D}a_\mu\int\mathcal{D}b_\mu
\exp\left\{i\int_{\mathbb{R}^{1,2}}d^3x\right.\nonumber\\
&\left[-\frac{1}{4}\left(\partial_{[\mu}a_{\nu]}-g\tilde{\chi}_{\mu\nu}\right)^2
+\,ea_\mu\epsilon^{\mu\nu\beta}\partial_\nu b_\beta\right.\nonumber\\
&\left.\left. +\,\alpha b_\mu\epsilon^{\mu\nu\beta}\partial_\nu b_\beta+2\pi b_\mu\Omega^\mu\right]\right\},
\label{eq21}
\end{align}
where we now sum over vortex configurations. Since we have a gaussian functional integral over the $b_\mu$ field, its integration is equivalent to substitute in (\ref{eq21}) the solution of its equation of motion. The equation of motion for the $b_\mu$ field is given by:
\begin{align}
\epsilon^{\mu\nu\beta}\partial_\nu\left(2\alpha b_\beta+ea_\beta\right)+2\pi\Omega^\mu=0.
\label{eq22}
\end{align}
By taking the divergence of the above equation, we obtain the constraint $\partial_\mu\Omega^\mu=0$, which tells us that the vortex density is a conserved current. We can solve this constraint locally by introducing a Chern-Kernel $\tilde{\lambda}_\mu$ for the vortex density:
\begin{align}
\Omega^\mu=\epsilon^{\mu\nu\beta}\partial_\nu\tilde{\lambda}_\beta,
\label{eq23}
\end{align}
where $\tilde{\lambda}_\mu=\frac{1}{2}\epsilon_{\mu\nu\beta}\lambda^{\nu\beta}$ is a delta distribution that localizes an arbitrary surface that has as boundary the closed line localized by the conserved vortex density $\Omega_\mu$. Using (\ref{eq23}), we obtain the general solution for the equation of motion (\ref{eq22}):
\begin{align}
b_\mu=-\frac{e}{2\alpha}a_\mu-\frac{\pi}{\alpha}\tilde{\lambda}_\mu+\partial_\mu\omega.
\label{eq24}
\end{align}
Substituting (\ref{eq24}) in the partition function (\ref{eq21}), we finally obtain the adequate definition of the MCS theory in the presence of external magnetic instantons \cite{mcsmon,artigao}:
\begin{align}
&Z_{\textrm{cond}}[j^3]=\sum_{\left\{\tilde{\lambda}_\mu\right\}}\int_{\textrm{G.F.}}\mathcal{D}a_\mu \exp\left\{i\int_{\mathbb{R}^{1,2}}d^3x\left[-\frac{1}{4}\left(\partial_{[\mu}a_{\nu]}\right.\right.\right.\nonumber\\
&\left.\left.\left.-\,g\tilde{\chi}_{\mu\nu}\right)^2-\frac{e^2}{4\alpha}\left(a_\mu+\frac{2\pi}{e}\tilde{\lambda}_\mu\right)
\epsilon^{\mu\nu\beta}\partial_\nu\left(a_\beta+\frac{2\pi}{e}\tilde{\lambda}_\beta\right)\right]\right\}.
\label{eq25}
\end{align}

Let us now take a break to analyze the issue of the magnetic Dirac brane symmetry in the low energy effective field theory (\ref{eq25}) for the electrically condensed regime within the $(2+1)$-dimensional layer. As discussed earlier in this section, the magnetic Dirac brane symmetry consists in the freedom of locating the magnetic Dirac strings at any place not occupied by the electric world-lines. However, in the electrically condensed regime, these electric world-lines proliferated occupying almost the whole space, except for the interior of the closed vortex lines, where the electric condensate vanishes. Therefore, in the electrically condensed regime, the only place allowed for the magnetic Dirac strings is the interior of closed magnetic vortices formally connected to the external magnetic instanton events. From (\ref{eq25}), we see that these vortices have a magnetic flux of $2\pi /e$. In this way, as explained in figure \ref{fig1}, provided the Dirac charge quantization condition is fulfilled, the closed magnetic vortices formally connected to the magnetic instantons have part of their flux canceled out by the flux inside the magnetic Dirac strings, giving rise to open magnetic vortices with a instanton-antinstanton pair in their boundaries.\\

\noindent
\begin{minipage}{\linewidth} % this is a non-floating enviroment that requires the caption package
\makebox[\linewidth]{\includegraphics[scale=0.5]{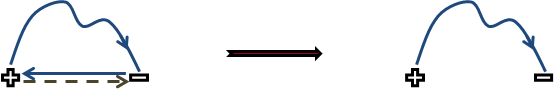}}
\captionof{figure}{\small{The magnetic flux inside the Dirac string (gray dashed arrow) is $g$, while the magnetic flux inside the closed vortex (blue solid line) is $2\pi /e$. The Dirac's veto implies that the only place allowed for the magnetic Dirac string in the electrically condensed regime is the interior of a closed magnetic vortex connected to a instanton-antinstanton pair. In this way, the Dirac charge quantization condition leads to a mutual cancellation between the magnetic flux inside the Dirac string and the magnetic flux inside the unphysical part of the closed vortex connected to the instanton-antinstanton pair. As the result of such a cancellation, there emerges a physical open magnetic vortex.}
\label{fig1}} % label inside the caption to assure correct reference in the body of the text
\end{minipage}
\vspace{6pt}

%\renewcommand{\figurename}{Figure} % this is a floating enviroment: the figure is not necessarily placed where you want
%\begin{figure}[h]
%       \centering
%       \includegraphics[scale=0.5]{flux-tube.png}
%       \caption{The magnetic flux inside the Dirac string (gray dashed arrow) is $g$, while the magnetic flux inside the closed vortex (blue solid line) is $2\pi /e$. The Dirac's veto implies that the only place allowed for the magnetic Dirac string in the electrically condensed regime is the interior of a closed magnetic vortex connected to a instanton-antinstanton pair. In this way, the Dirac charge quantization condition leads to a mutual cancellation between the magnetic flux inside the Dirac string and the magnetic flux inside the unphysical part of the closed vortex connected to the instanton-antinstanton pair. As the result of such a cancellation, there emerges a physical open magnetic vortex.
%\label{fig1}} % label inside the caption to assure correct reference in the body of the text
%\end{figure}

Notice that due the cancellation scheme depicted in figure \ref{fig1}, we can freely deform the unphysical magnetic Dirac strings in the electrically condensed regime, as long as the unphysical part of the closed magnetic vortices that wipe out the Dirac strings transforms accordingly. In fact, this is necessary to preserve the Dirac's veto in the condensed regime. Therefore, in the electrically condensed phase, the magnetic Dirac brane symmetry (\ref{eq17}) becomes \cite{jt-cho-su2}:
\begin{align}
\tilde{\chi}_{\mu\nu}&\mapsto\tilde{\chi}_{\mu\nu}+\partial_{[\mu}\tilde{\sigma}_{\nu]},\nonumber\\
a_\mu&\mapsto a_\mu+g\tilde{\sigma}_{\mu},\nonumber\\
\tilde{\lambda}_\mu&\mapsto \tilde{\lambda}_\mu-\tilde{\sigma}_{\mu}.
\label{eq26}
\end{align}
From this, we see that the local magnetic Dirac brane symmetry is preserved in the MCS theory in the presence of external magnetic instantons (\ref{eq25}) derived here via GJTA, in consonance with Elitzur's theorem \cite{elitzur}.

Before proceeding to discuss the confinement of the magnetic instantons featured in (\ref{eq25}), let us firstly analyze the issue of the quantization of the topological mass of the gauge field $a_\mu$ due to the presence of the magnetic instantons. As reviewed in \cite{dunne}, the coefficient of the CS term, $a_\mu\epsilon^{\mu\nu\beta}\partial_\nu a_\beta$, gives the physical propagating pole of the MCS theory. In this manner, we can immediately read off from (\ref{eq25}) the gauge invariant topological mass acquired by the $a_\mu$ field: $m_{\textrm{top}}=e^2/2\alpha$. From (\ref{eq23}) and (\ref{eq25}), we see that the gauge field couples minimally to the magnetic vortex density $\Omega_\mu$ and that the coefficient of this minimal coupling term is given by $e\pi/\alpha$. Since the magnetic vortex density couples minimally to the gauge field, we see that it acquires an electric charge $e\pi/\alpha$, developing a dyonic character. Due to the Dirac charge quantization condition, any electric charge in the system must be quantized in integer multiples of $2\pi/g$ and, therefore, we get the following identity for the electric charge of the vortices:\begin{align}
\frac{e\pi}{\alpha}=\frac{2\pi n}{g},\,n\in\mathbb{Z}.
\label{eq27}
\end{align}
By making the reasonable assumption that the electric charges $e$ that condensed has minimal strength $e=2\pi/g$, we obtain from (\ref{eq27}):
\begin{align}
\alpha=\frac{eg\pi}{2\pi n}=\frac{2\pi^2}{2\pi n}=\frac{\pi}{n},\,n\in\mathbb{Z},
\label{eq28}
\end{align}
thus fixing the possible values of the GJTA parameter $\alpha$, as promised before. With this result in hands, we obtain the following quantization rule for the topological mass of the gauge field due to the presence of magnetic instantons \cite{artigao}:
\begin{align}
m_{\textrm{top}}=\frac{e^2}{2\alpha}=\frac{(2\pi/g)^2}{2(\pi/n)}=\frac{2\pi n}{g^2},\,n\in\mathbb{Z},
\label{eq29}
\end{align}
which agrees with the quantization rule found originally in \cite{ht}.

In order to adequately address the issue of the magnetic instanton confinement in the MCS theory, let us rewrite the partition function (\ref{eq25}) in terms of observable Dirac brane invariants \cite{artigao}. We begin by defining the following Dirac brane invariant gauge field:
\begin{align}
\bar{a}_\mu:=a_\mu+\frac{2\pi}{e}\tilde{\lambda}_\mu,
\label{eq30}
\end{align}
which does not change under the magnetic Dirac brane tranformation in the electrically condensed phase (\ref{eq26}). We then have $a_\mu=\bar{a}_\mu-\frac{2\pi}{e}\tilde{\lambda}_\mu$, such that the partition function (\ref{eq25}) can be rewritten in the following form \cite{artigao}:
\begin{align}
Z_{\textrm{cond}}[j^3]=&\sum_{\left\{\tilde{\lambda}_\mu\right\}}\int_{\textrm{G.F.}}\mathcal{D}\bar{a}_\mu \exp\left\{i\int_{\mathbb{R}^{1,2}}d^3x\left[-\frac{1}{4}\left(\partial_{[\mu}\bar{a}_{\nu]}\right.\right.\right.\nonumber\\
&\left.\left.\left.-\,g\tilde{L}_{\mu\nu}\right)^2-\frac{m_{\textrm{top}}}{2}\,\bar{a}_\mu
\epsilon^{\mu\nu\beta}\partial_\nu\bar{a}_\beta\right]\right\},
\label{eq31}
\end{align}
where we made use of the Dirac charge quantization condition to define the magnetic Dirac brane invariant distribution:
\begin{align}
\tilde{L}_{\mu\nu}:=\tilde{\chi}_{\mu\nu}+\partial_{[\mu}\tilde{\lambda}_{\nu]},
\label{eq32}
\end{align}
which does not change under the transformation (\ref{eq26}). From (\ref{eq32}), we see that the scalar instanton current is written in terms of the Dirac brane invariant as:
\begin{align}
j^3=\frac{1}{2}\varepsilon^{3\mu\nu\beta}\partial_\mu\tilde{\chi}_{\nu\beta}\equiv \frac{1}{2}\epsilon^{\mu\nu\beta} \partial_\mu\tilde{\chi}_{\nu\beta}&=\frac{1}{2}\epsilon^{\mu\nu\beta}\partial_\mu\tilde{L}_{\nu\beta}\nonumber\\
&=\partial_\mu L^\mu,
\label{eq33}
\end{align}
where $L_\mu=\frac{1}{2}\epsilon_{\mu\nu\beta}\tilde{L}^{\nu\beta}$.

The partition function (\ref{eq31}) is entirely written in terms of magnetic Dirac brane invariants, such that the magnetic Dirac brane symmetry (\ref{eq26}) is hidden realized in the interior of these invariants. This hiding of the Dirac brane symmetry was advocated by us in \cite{artigao} as a general signature of the confinement phenomenon in physical systems characterized by condensates of topological currents. As discussed before, in the presence of external magnetic instantons, there are both, closed vortices connected and disconnected from the instantons. As explained in figure \ref{fig1}, the closed vortices connected to the instantons give rise to open vortices with instanton-antinstanton pairs in their boundaries. These open vortices are the magnetic confining flux tubes. If we now consider the system within the $(2+1)$-dimensional layer in a particular state with no disconnected vortices from the magnetic instantons, and integrate out the gauge field, the partition function (\ref{eq31}) can be shown to reduce to \cite{mcsmon,artigao}:
\begin{align}
&Z_{\textrm{cond}}^{(\textrm{open})}[j^3]=\sum_{\left\{L_\mu\right\}}^{(\textrm{open})} \exp\left\{i\int_{\mathbb{R}^{1,2}}d^3x\right.\nonumber\\
&\left.\frac{g^2}{2}L_\mu\left(\frac{m_{\textrm{top}}\epsilon^{\mu\beta\nu}\partial_\beta +\partial^\mu\partial^\nu-m_{\textrm{top}}^2\eta^{\mu\nu}}{\partial^2+m_{\textrm{top}}^2}\right)L_\nu\right\},
\label{eq34}
\end{align}
where the sum over geometric configurations is now performed only over the open vortices. The effective action present in the Boltzmann factor in (\ref{eq34}) agrees with the action originally obtained in \cite{diamantini}. At asymptotic instanton-antinstanton separations, we expect that the relevant contribution in the sum over all possible shapes of the confining flux tubes in the partition function (\ref{eq34}) is given by the open vortex configuration minimizing the energy of the system, that is presumably a straight line giving the minimal interval between two instanton and antinstanton events in $\mathbb{R}^{1,2}$. The third term in the effective action in the Boltzmann factor in (\ref{eq34}) is dominant at asymptotic distances and its contribution for the effective interaction potential associated to an instanton-antinstanton configuration connected by a straight flux tube is given by (see \cite{artigao} for the detailed evaluation):
\begin{align}
V_{\textrm{eff}}(R)\stackrel{R\rightarrow\infty}{\sim}\frac{g^2 m_{\textrm{top}}^2}{8\pi}\ln
\left(\frac{m_{\textrm{top}}^2+M^2}{m_{\textrm{top}}^2}\right)R,
\label{eq35}
\end{align}
where $R$ is the instanton-antinstanton separation and the coefficient of the confining linear potential (\ref{eq35}) is the so-called string tension. The parameter $M$ featured in (\ref{eq35}) is a physical ultraviolet cutoff corresponding here to the inverse of the coherence length of the electric condensate within the $(2+1)$-dimensional layer. This coherence length is also the thickness of the magnetic confining flux tubes.

Therefore, we see that the presence of fermions drastically modifies the low energy physics within the $(2+1)$-dimensional layer of the dual Josephson junction setup considered here. In the absence of fermions, there is a weak electric charge confinement within the layer \cite{tetradis}, as reviewed in section \ref{sec2}. However, if there are fermions in the dual Josephson junction, their net effect, which we computed in this section by making use of the GJTA, is to destroy the weak electric charge confinement and to promote the confinement of the magnetic instantons within the $(2+1)$-dimensional layer.

\section{Concluding remarks}
\label{sec4}

In this work, we analyzed an extension of the model proposed in \cite{tetradis}, where a dual Josephson junction comprising a $(2+1)$-dimensional non-superconducting layer sandwiched between two dual superconducting regions in $(3+1)$-dimensions works as a model for localizing a $U(1)$ gauge field within the layer. This localization mechanism is provided by the fact that the photons within the layer cannot escape from it to propagate far into the dual superconducting bulk, since in the interior of the bulk they would acquire a mass due to the dual Higgs mechanism. The photon propagation, therefore, remains effectively localized within the layer. When a phase difference between the wave functions describing the monopole condensate constituting the dual superconducting bulk below and above the non-superconducting layer is taken into consideration, there is a flow of monopole currents through the layer, which appears within it as a gas of magnetic instanton events. In the absence of fermions, this magnetic instanton gas promotes a weak electric charge confinement within the layer, which is only noticeable at very long distances.

The extension we considered in the present work, was the inclusion of fermions in this dual Josephson junction setup. Due to the fact that the dual superconducting bulk features a dual Meissner effect, which tends to expel all the electric fields from the interior of the dual superconducting medium, causing the confinement of electric charges, unconfined fermions would be localized within the $(2+1)$-dimensional non-superconducting layer, where their quantum fluctuations induce a Chern-Simons term. We derived this effect by making use of the generalized Julia-Toulouse approach for condensation of topological currents, interpreting the induction of the Chern-Simons term as an electric condensation process. As the result, in the presence of fermions, the weak electric charge confinement within the layer is destroyed and the magnetic instantons become confined due to the presence of the electric condensate effectively described by the Chern-Simons term.

\acknowledgments

We thank Conselho Nacional de Desenvolvimento Cient\'ifico e Tecnol\'ogico (CNPq) for financial support.

% referências ordenadas pela ordem com que são citadas no artigo

\end{document}